\documentclass[aps, prd, twocolumn, showpacs, showkeys, superscriptaddress, preprintnumbers, floatfix]{revtex4}
\usepackage{multirow}
\usepackage{slashed}
\usepackage{graphicx}
\usepackage{amsmath}
\usepackage{bm}
\usepackage{yhmath}
\usepackage{hyperref}
\usepackage{subfigure}
\usepackage[T1]{fontenc}
\usepackage{color}
\usepackage{cases}
\usepackage{subfigure}
\usepackage{dcolumn, booktabs, bm}
\usepackage{amsfonts, amssymb, stmaryrd, latexsym, amsmath}
\usepackage{textcomp}
\usepackage{amsthm}
\usepackage{mathrsfs}

\usepackage{array}

\newcounter{RomanNumber}

\newcommand{\lyxmathsym}[1]{\ifmmode\begingroup\def\b@ld{bold}
	\text{\ifx\math@version\b@ld\bfseries\fi#1}\endgroup\else#1\fi}

\renewcommand{\arraystretch}{1.9}
\usepackage{braket}
\allowdisplaybreaks

\begin{document}
	\title{Where is the next pentaquark state?}
	
	\author{Hao-Song Li}
	\email{haosongli@nwu.edu.cn}
	\affiliation{School of Physics, Northwest University, Xian 710127, China}
	\affiliation{Shaanxi Key Laboratory for Theoretical Physics Frontiers, Xian 710127, China}
	\affiliation{Peng Huanwu Center for Fundamental Theory, Xian 710127, China}
	\author{Tao Li}
	\email{tao@stumail.nwu.edu.cn}
	\affiliation{School of Physics, Northwest University, Xian 710127, China}
	\begin{abstract}
	The LHCb Collaboration has observed two types of hidden-charm pentaquark states $P_{\psi}^N$ and $P_{\psi s}^{\Lambda}$ since 2015. In this work, we predict two other types of hidden-charm pentaquark states $P_{\psi s}^{\Sigma}(4367)$ and $P_{\psi ss}^{N}(4379)$ within the framework of heavy pentaquark chiral perturbation theory. We suggest the LHCb Collaboration to observe $P_{\psi ss}^{N}(4379)^{-}$ and $P_{\psi ss}^{N}(4379)^0$ with $J^P=\frac{1}{2}^{-}$ in	the $J/\psi \Xi$ spectrum through amplitude analyses of $\Omega_b^- \to J/\psi \Xi^0 K^-$ decays and $B^- \to J/\psi \Xi^- \bar{\Lambda}$ decays.
	\end{abstract}
	\maketitle
	\thispagestyle{empty}

	Hadron physics is an important research area in particle physics, while exploring the structure of different hadrons is the frontier of particle physics. With the continuous progress of experimental technology, particularly the LHCb detector, a large number of hadrons have been observed. Among them, the hidden-charm pentaquark state with unique structure has attracted much attention. In 2015, the LHCb Collaboration first observed two hidden-charm pentaquark states $P_{\psi}^N(4380)$ and $P_{\psi}^N(4450)$~\cite{LHCb:2015yax}. In 2019, with the update of the experimental data, $P_{\psi}^N(4450)$ appeared to be
	split into two narrower structures $P_{\psi}^N(4440)$ and $P_{\psi}^N(4457)$, and a third peak  $P_{\psi}^N(4312)$ showed up~\cite{LHCb:2019kea}. In 2020, the hidden-charm strange pentaquark state $P_{\psi{}s}^{\Lambda}(4459)^{0}$ was observed by the LHCb Collaboration~\cite{LHCb:2020jpq}. Most recently, a new structure $P_{\psi{}s}^{\Lambda}(4338)^{0}$ has been observed as well, whose spin-parity is $J^P = \frac{1}{2}^-$ with 90\% probability~\cite{LHCb:2022ogu}. We have listed the masses, widths, observed decay channels and thresholds of the pentaquark states in Table ~\ref{tab_1}.
	
	The pentaquark states are exotic hadrons consist of four quarks and one antiquark beyond the conventional quark model. The internal structure and properties of the pentaquark states have not yet been fully understood, various theoretical frameworks have been proposed to explain the pentaquark states~\cite{Chen:2016qju,Liu:2013waa,Yuan:2018inv,Olsen:2017bmm,Guo:2017jvc,Meng:2022ozq,Xiao:2019gjd,Liu:2020hcv,Peng:2020hql,Zhu:2021lhd,Du:2021bgb,	Chen:2021tip,Yang:2021pio,Deng:2022vkv,Wang:2019got,Cheng:2019obk,Weng:2019ynv,Zhu:2019iwm,Pimikov:2019dyr,Liu:2024uxn}. The prevailing picture
	is that the pentaquark states are primarily hadronic molecules, since the pentaquark states observed by LHCb are located near the mass thresholds of various charmed mesons and baryons~\cite{Liu:2024dlc}. 
	Specifically, $P_{\psi}^N(4312)$, $P_{\psi}^N(4440)$ and $P_{\psi}^N(4457)$ are close to the $\bar{D}\Sigma_c$ and $\bar{D}^*\Sigma_c$ thresholds~\cite{Wu:2010jy,Wang:2011rga,Yang:2011wz,Yuan:2012wz,Wu:2012md,Xiao:2013yca,Uchino:2015uha,Karliner:2015ina,Liu:2018zzu,Sakai:2019qph,Du:2021fmf,Peng:2022iez}, $P_{\psi s}^{\Lambda}(4380)$ is close to the $\bar{D}^*\Xi_c$ threshold, and $P_{\psi s}^{\Lambda}(4459)$ is close to the $\bar{D}\Xi_c^{'}$ threshold~\cite{Feijoo:2022rxf,Wang:2023eng}. In addition to the hadronic molecule picture, there are many other interpretations, for instance, $P_{\psi}^N(4312)$ is explained as a virtual state in Ref.~\cite{Fernandez-Ramirez:2019koa} or a bound state of $\chi_{c0}$ with nucleon in Ref.~\cite{Eides:2019tgv}, $P_{\psi}^N(4457)$ is explained as cusp of $\Sigma_c\bar{D}^*$~\cite{Burns:2022uiv}. In this paper we work in the context of the hadronic molecular picture with chiral perturbation theory.
	\renewcommand{\arraystretch}{2.8}	
	\begin{table}[htp]
		\centering
		\caption{Data of hidden-charm pentaquark states}
		\vspace{0.1em}
		\label{tab_1} 
		\renewcommand{\arraystretch}{2.5}
		\resizebox{82mm}{!}{
			\begin{tabular}{ c|c|c|c|c}
				\toprule[1.0pt]
				\toprule[1.0pt]
				Pentaquarks & Masses$(\mbox{MeV})$ & Widths$(\mbox{MeV})$&Channels  & Near thresholds
				\\
				\hline
				$P^{N}_{\psi}(4380)^{^{+}}$ & $4380\pm8\pm29$          &       $215\pm18\pm86$           & $\Lambda_b^0 \to J/\psi pK^-$&$\Sigma_c^*\bar{D}$\\ \hline
				$P^{N}_{\psi}(4312)^{^{+}}$ & ~~$4311.9\pm0.7^{~+6.8}_{~-0.6}$~~  &  $9.8\pm2.7^{~+3.7}_{~-4.5}$    & $\Lambda_b^0 \to J/\psi pK^-$&$\Sigma_c\bar{D}$ \\ \hline
				$P^{N}_{\psi}(4440)^{^{+}}$ & $4440.3\pm1.3^{~+4.1}_{~-4.7}$   &  $20.6\pm4.9^{~+8.7}_{~-10.1}$  &$\Lambda_b^0 \to J/\psi pK^-$&$\Sigma_c\bar{D}^*$ \\ \hline
				$P^{N}_{\psi}(4457)^{^{+}}$ & $4457.3\pm0.6^{~+4.1}_{~-1.7}$   &  $6.4\pm2.0^{~+5.7}_{~-1.9}$    & $\Lambda_b^0 \to J/\psi pK^-$ &$\Sigma_c\bar{D}^*$\\ \hline
				
				$P_{\psi{}s}^{\Lambda}(4459)^{0}$& $4458.8\pm2.9^{~+4.7}_{~-1.1}$   &  $17.3\pm6.5^{~+8.0}_{~-5.7}$   &$\Xi_b^- \to J/\psi \Lambda K^-$&$\Xi_c\bar{D}^*$\\ \hline
				$P_{\psi{}s}^{\Lambda}(4338)^{0}$ &    $4338.2 \pm 0.7 \pm 0.4$   &  $7.0 \pm 1.2 \pm 1.3$    & $B^-\to J/\psi \Lambda \bar{p}$&$\Xi_c\bar{D}$($\frac{1}{2}^-$)
				\\
				\toprule[1.0pt]
				\toprule[1.0pt]
		\end{tabular}}
	\end{table}	
	
	Chiral perturbation theory provides a systematic, perturbatively computable theoretical framework for the study of hadron physics in the low-energy region~\cite{Weinberg:1978kz,Gasser:1983yg,Gasser:1984gg,Bernard:1995dp,Scherer:2002tk}. It enables effective theoretical analysis and prediction of the low-energy properties of the hadron without directly resolving the complex quantum chromodynamics. For the special hadronic structure of pentaquark states, specific theories are needed to describe their properties more accurately, thus we develop the heavy pentaquark chiral perturbation theory (HPChPT)~\cite{Li:2024jlq}. In the low-energy limit, the strong interactions have an approximate chiral symmetry, but this symmetry is broken, and HPChPT exploits this symmetry-breaking property to construct the theory. Similar to baryon chiral perturbation theory, HPChPT can also perform systematic perturbation calculations to approximate the real results order-by-order.
	
 	Since the LHCb Collaboration has observed two types of hidden-charm pentaquark states $P_{\psi}^N$ and $P_{\psi s}^{\Lambda}$, in this work, we predict two other types of hidden-charm pentaquark states $P_{\psi s}^{\Sigma}$ and $P_{\psi ss}^{N}$ and investigate the chiral corrections to the masses of spin-$\frac{1}{2}$ octet hidden-charm molecular pentaquark states with HPChPT. We adopt a framework established in Ref. \cite{Li:2024jlq}. As the hidden-charm molecular pentaquark states are composed of the corresponding singly charmed baryons and anti-charmed mesons, considering the direct product $3\otimes3\otimes3 = 1\oplus8_1\oplus8_2\oplus10$, the octet hidden-charm molecular pentaquark $P_n$ reads
\begin{equation}
	\setlength{\arraycolsep}{0.1pt}
	P_n=
	\left(		
	\begin{array}{ccc}
		\frac{1}{\sqrt{2}}{{}P_{\psi s}^{\Sigma}}^{0}+\frac{1}{\sqrt{6}}{{}P_{\psi s}^{\Lambda}}^{0}
		&{{}P_{\psi s}^{\Sigma}}^{+}
		&{{}P_{\psi}^{N}}^{+}
		\\
		{{}P_{\psi s}^{\Sigma}}^{-}
		&-\frac{1}{\sqrt{2}}{{}P_{\psi s}^{\Sigma}}^{0}+\frac{1}{\sqrt{6}}{{}P_{\psi s}^{\Lambda}}^{0}
		&{{}P_{\psi}^{N}}^{0}
		\\
		{{}P_{\psi ss}^{N}}^{-}
		&{P_{\psi ss}^{N}}^{0}
		&-\frac{2}{\sqrt{6}}{{}P_{\psi s}^{\Lambda}}^{0}
		\nonumber\\
	\end{array}
	\right),
\end{equation}
	The flavor wave functions of the octet hidden-charm molecular pentaquark under $SU(3)$ symmetry are shown in Table~\ref{tab111} (details in Ref.~\cite{Li:2024wxr}).
		\begin{table}[h!]
		\renewcommand{\arraystretch}{2}
		\centering
		\caption{$8_{1}$ and $8_{2}$ flavor wave functions}
		\label{tab111} 		
		\begin{tabular}{c|c|c}
			\toprule[1.0pt]
			\toprule[1.0pt]
			States &  Flavor & Wave functions
			\\
			\hline
			{${P_{\psi}^{N+}}$}
			&	$8_{1}$
			&	$-\sqrt{\frac{1}{3}}\Sigma_c^{+}\bar{D}^{(*)0}+\sqrt{\frac{2}{3}}\Sigma_{c}^{++}D^{(*)-}$
			\\

			&	$8_{2}$
			&	$\Lambda_{c}^{+}\Bar{D}^{(*)0}$
			\\
			\hline
			
			{${P_{\psi}^{N0}}$}
			&	$8_{1}$
			&	$\sqrt{\frac{1}{3}}\Sigma_c^{+}D^{(*)-}-\sqrt{\frac{2}{3}}\Sigma_{c}^{0}\bar{D}^{(*)0}$
			\\

			&	$8_{2}$
			&	$\Lambda_{c}^{+}D^{(*)-}$
			\\
			\hline
			
			{${P_{\psi s}^{\Sigma+}}$}
			&	$8_{1}$
			&	$\sqrt{\frac{1}{3}}\Xi_c^{\prime+}\bar{D}^{(*)0}-\sqrt{\frac{2}{3}}\Sigma_{c}^{++}D_s^{(*)-}$
			\\

			&	$8_{2}$
			&	$\Xi_{c}^{+}\Bar{D}^{(*)0}$
			\\
			\hline
			
			{${P_{\psi s}^{\Sigma0}}$}
			&	$8_{1}$
			&	$\sqrt{\frac{1}{6}}\Xi_c^{\prime+}D^{(*)-}+\sqrt{\frac{1}{6}}\Xi_{c}^{\prime0}\bar{D}^{(*)0}-\sqrt{\frac{2}{3}}\Sigma_{c}^{+}D_s^{(*)-}$
			\\

			&	$8_{2}$
			&	$\sqrt{\frac{1}{2}}\Xi_c^{+}D^{(*)-} + \sqrt{\frac{1}{2}}\Xi_{c}^{0}\bar{D}^{(*)0} $
			\\
			\hline
			
			{${P_{\psi s}^{\Lambda0}}$}
			&	$8_{1}$
			&	$\sqrt{\frac{1}{2}}\Xi_c^{\prime+}D^{(*)-} - \sqrt{\frac{1}{2}}\Xi_{c}^{\prime0}\bar{D}^{(*)0} $
			
			\\

			&	$8_{2}$
			&	$\sqrt{\frac{1}{6}}\Xi_c^{+}D^{(*)-}-\sqrt{\frac{1}{6}}\Xi_{c}^{0}\bar{D}^{(*)0}-\sqrt{\frac{2}{3}}\Lambda_{c}^{+}D_s^{(*)-}$
			\\
			\hline
			
			{${P_{\psi s}^{\Sigma-}}$}
			&	$8_{1}$
			&	$\sqrt{\frac{1}{3}}\Xi_c^{\prime0}D^{(*)-}-\sqrt{\frac{2}{3}}\Sigma_{c}^{0}D_s^{(*)-}$
			\\

			&	$8_{2}$
			&	$\Xi_{c}^{0}D^{(*)-}$
			\\
			\hline
			
			{${P_{\psi ss}^{N0}}$}
			&	$8_{1}$
			&	$\sqrt{\frac{1}{3}}\Xi_c^{\prime+}D_{s}^{(*)-}-\sqrt{\frac{2}{3}}\Omega_{c}^{0}\bar{D}^{(*)0}$
			\\
			
			&	$8_{2}$
			&	$\Xi_{c}^{+}D_{s}^{(*)-}$
			\\
			\hline
			{${P_{\psi ss}^{N-}}$}
			&	$8_{1}$
			&	$\sqrt{\frac{1}{3}}\Xi_c^{\prime0}D_{s}^{(*)-}-\sqrt{\frac{2}{3}}\Omega_{c}^{0}D^{(*)-}$
			\\

			&	$8_{2}$
			&	$\Xi_{c}^{0}D_{s}^{(*)-}$
			\\
			\toprule[1.0pt]
			\toprule[1.0pt]
		\end{tabular}	
	\end{table}

	

 To calculate the masses of spin-$\frac{1}{2}$ octet pentaquark states, we construct the  Lagrangian contributing to the octet pentaquark mass at the leading order,
	\begin{eqnarray}
		\label{tree1}
		\mathcal{L}^{(2)}_1=b_{1} {\rm Tr}(\bar{P_1}\{\chi_+,P_1\})+b_{2}{\rm Tr}(\bar{P_1}[\chi_+,P_1]),  \label{Eq:Mass1}
			\end{eqnarray}
			\begin{eqnarray}
		\mathcal{L}^{(2)}_2=c_{1} {\rm Tr}(\bar{P_2}\{\chi_+,P_2\})+c_{2}{\rm Tr}(\bar{P_2}[\chi_+,P_2]), \label{Eq:Mass2}
	\end{eqnarray}
where $P_{1,2}$ represents the $8_{1,2}$ flavor pentaquark state respectively, $b_{{1,2}}$ and $c_{{1,2}}$ are low-energy constants (LECs), $\chi_{+}$ reads
	\begin{eqnarray}
\chi_{+}=u^\dag\chi u^\dag+ u\chi^\dag u,
	\end{eqnarray}
	where $\chi=diag(M_{\pi}^2, M_{\pi}^2, 2M_K^2-M_{\pi}^2)$, $u=\exp(i\phi/2F_{\varphi})$, with the subscript $\varphi=\pi, K, \eta$. We use the pseudoscalar meson decay constants $F_{\pi}\approx$ 92.4 MeV, $F_{K}\approx$ 113 MeV and
	$F_{\eta}\approx$ 116 MeV, $M_{\pi}=0.140$ GeV, $M_{K}=0.494$ GeV and	$M_{\eta}=0.550$ GeV. The pseudoscalar meson field $\phi$ reads
	\begin{equation}
		\phi=\left(\begin{array}{ccc}
			\pi^{0}+\frac{1}{\sqrt{3}}\eta & \sqrt{2}\pi^{+} & \sqrt{2}K^{+}\\
			\sqrt{2}\pi^{-} & -\pi^{0}+\frac{1}{\sqrt{3}}\eta & \sqrt{2}K^{0}\\
			\sqrt{2}K^{-} & \sqrt{2}\bar{K}^{0} & -\frac{2}{\sqrt{3}}\eta
		\end{array}\right).
	\end{equation}

	We also need the interaction Lagrangians of $8_{1}$ and $8_{2}$ flavor pentaquark states contributing to their masses on the loop level,
	\begin{equation}
		\mathcal{L}_{\rm
			int1}=2g_{1}{\rm Tr}(\bar{P}_1 S_{\mu}\{u^{\mu},P_1\}) +2f_{1}{\rm Tr}(\bar{P}_1S_{\mu}[u^{\mu},P_1]),
		\label{Eq:baryon21}
	\end{equation}
	\begin{equation}
		\mathcal{L}_{\rm
			int2}=2g_{4}{\rm Tr}(\bar{P}_2 S_{\mu}\{u^{\mu},P_2\}) +2f_{4}{\rm Tr}(\bar{P}_2S_{\mu}[u^{\mu},P_2]),
		\label{Eq:baryon22}
	\end{equation}
	where $S_{\mu} $ is a covariant spin operator,  $g_{1,4}$ and $f_{1,4}$ are $\phi PP$ coupling constants. The axial vector field $u_{\mu}$ is defined as~\cite{Scherer:2002tk}
	\begin{eqnarray}
		u_{\mu}&=&\frac{1}{2}i\left[u^{\dagger}\partial_{\mu}u-u\partial_{\mu}u^{\dagger}\right].
	\end{eqnarray} 
	
	In the framework of HPChPT, one decomposes the four-momentum $p_{\mu}$
	of a low-energy pentaquark state into $m_0v_{\mu}$ and a residual momentum $k_{\mu}$,
	\begin{eqnarray}
		p_{\mu}=m_0v_{\mu}+k_{\mu},
	\end{eqnarray}
	where $m_0$ is the bare mass of the hidden-charm octet pentaquark state, $v_{\mu}=(1,\vec{0})$ is the velocity of the pentaquark state satisfing $v\cdot k 	\ll m_0$.
	
\begin{figure}[tbh]
	\includegraphics[width=1.1\hsize]{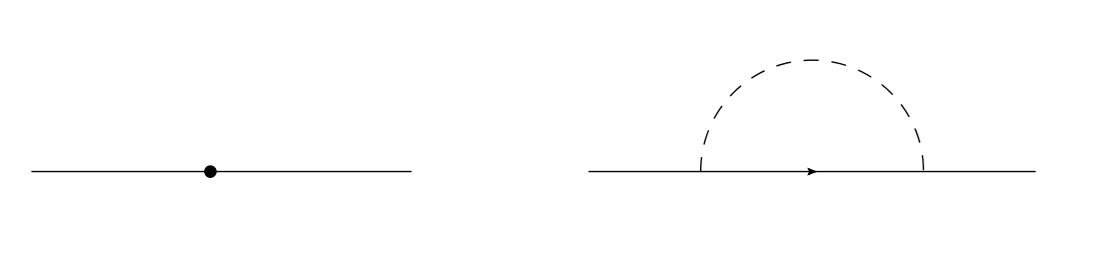}
	\caption{Feynman diagrams contributing to the self-energy of hidden-charm pentaquark states. The dot represent $\mathcal{O}(p^{2})$ coupling. The solid line and the dashed line represent the octet hidden-charm pentaquark state and the pseudoscalar meson respectively.}
	\label{fig:tree}
\end{figure}

	The full form propagator of the pentaquark state reads,
	\begin{eqnarray}
		G&=&\frac{i}{v\cdot p-m_0-\Sigma_{P}(p)},
		\label{9}
	\end{eqnarray}	
		where $\Sigma_P(p)$ denotes the high order contributions to the self-energy in Fig.~\ref{fig:tree}. To facilitate future calculations, we need two scalar functions $\eta = v\cdot p-m$ and $\xi = (p - mv)^{2}$, and Eq.~(\ref{9}) is written as
			\begin{eqnarray}
				G&=&\frac{i}{v\cdot p-m_0-\Sigma_P(\eta,\xi)}\nonumber\\
				&=&\frac{iZ_P}{v\cdot
					p-m-Z_P\tilde{\Sigma}_P(\eta,\xi)},\label{self}
					\end{eqnarray}	
where $Z_P$ is the reformation constant,
\begin{equation}
	Z_P=\frac{1}{1-\Sigma_P^\prime(0,0)},
\end{equation}
with
\begin{equation}
	\Sigma_P^\prime(0,0)=\frac{\partial \Sigma_P(\eta,\xi)}{\partial
		\eta}\big|_{(\eta,\xi)=(0,0)}.
\end{equation}

The complete propagator has a pole at $p=mv$, which includes both the mass-shell conditions $p^2=m^2$ and $v\cdot p=m$, $m$ is the physical mass of the octet pentaquark state. Thus, the masses of the pentaquark states are
\begin{equation}
	m_P=m_0+\Sigma_P(0,0). \label{eqmSum}
\end{equation}


The chiral contribution to the pentaquark self-energy up to next-to-leading order (NLO) includes two parts in Fig.~\ref{fig:tree}. The tree level self-energy terms come from Eq. (\ref{Eq:Mass1}) and Eq. (\ref{Eq:Mass2}). The loop level self-energy terms can be written as
\begin{eqnarray}
	\Sigma^{(\rm loop)}_{P_{1,2},\varphi}&=&-{C^{(\rm loop)}_{P_{1,2}\varphi}}\frac{1}{(4\pi F_{\varphi})^2}\left\{\frac{v\cdot
		k}{4}\right.\Big(\left[3M_{\varphi}^2-2(v\cdot
	k)^2\right]\nonumber\\
	&&\left[R+\ln\left(\frac{M_{\varphi}^2}{\Lambda_{\chi}^2}\right)\right] 
	-2\left[M_{\varphi}^2-(v\cdot
	k)^2\right]\Big)\nonumber\\
	&&+\left[M_{\varphi}^2-(v\cdot
	k)^2\right]^{3/2}\left.\arccos\left(-\frac{v\cdot
		k}{M_{\varphi}}\right)\right\} \nonumber\\
			&&+{C^{(\rm loop)}_{P_{1,2}\varphi}}\frac{1}{(4\pi F_{\varphi})^2}(v \cdot k)\Big[\frac{M_{\varphi}^2}{2}-\frac{(v \cdot k)^2}{2}\Big],
\end{eqnarray}
where the coefficients $C^{(\rm loop)}_{P_{1,2}\varphi}$ are given in Table~\ref{tab_3}, $\varphi=\pi, K, \eta$, $\Lambda_{\chi}$ is the cutoff energy scale, $R=\frac{2}{d-4}-[\ln(4\pi)+\varGamma{'}(1)+1]$, $d$ is the space-time dimension.

	\begin{table}[htp]
	\centering
	\caption{Coefficients ${C^{(\rm loop)}_{P_{1,2}\varphi}}$}
	\vspace{0.1em}
	\label{tab_3} 
\renewcommand{\arraystretch}{2.5}
	\resizebox{82mm}{!}{
			\begin{tabular}{ c|c|c|c|c}
			\toprule[1.0pt]
			\toprule[1.0pt]
			States & ${P_{\psi}^{N}}$&${P_{\psi s}^{\Sigma}}$&${P_{\psi s}^{\Lambda0}}$&${P_{\psi ss}^{N}}$\\ \hline
			$C^{(\rm loop)}_{P_1\pi}$&$3(f_1+g_1)^2$&$\frac{4}{3}(6f_1^2+g_1^2)$&$4g_1^2$&$3(f_1-g_1)^2$ \\ \hline
			$C^{(\rm loop)}_{P_1K}$&$\frac{2}{3}(9f_1^2-6f_1g_1+5g_1^2)$&$4(f_1^2+g_1^2)$&$\frac{4}{3}(9f_1^2+g_1^2)$&$\frac{2}{3}(9f_1^2+6f_1g_1+5g_1^2)$\\\hline
			$C^{(\rm loop)}_{P_1\eta}$&$\frac{1}{3}(g_1-3f_1)^2$&$\frac{4}{3}g_1^2$&$\frac{4}{3}g_1^2$&$\frac{1}{3}(3f_1+g_1)^2$\\ \hline 
					$C^{(\rm loop)}_{P_2\pi}$&$3(f_4+g_4)^2$&$\frac{4}{3}(6f_4^2+g_4^2)$&$4g_4^2$&$3(f_4-g_4)^2$ \\ \hline
				$C^{(\rm loop)}_{P_2K}$&$\frac{2}{3}(9f_4^2-6f_4g_4+5g_4^2)$&$4(f_4^2+g_4^2)$&$\frac{4}{3}(9f_4^2+g_4^2)$&$\frac{2}{3}(9f_4^2+6f_4g_4+5g_4^2)$\\\hline
			$C^{(\rm loop)}_{P_2\eta}$&$\frac{1}{3}(g_4-3f_4)^2$&$\frac{4}{3}g_4^2$&$\frac{4}{3}g_4^2$&$\frac{1}{3}(3f_4+g_4)^2$\\
			\toprule[1.0pt]
			\toprule[1.0pt]
				
		\end{tabular}}
	\end{table}	
To NLO, the physical masses of $8_1$ flavor hidden-charm pentaquark states read
	\begin{eqnarray}
		\label{equ:20}
	m_{P_{\psi}^{N}}&=&m_0+4b_{2}(M_K^2-M_\pi^2)-4b_{1}M_K^2\nonumber\\
	&&-\frac{3(f_1+g_1)^2 M_{\pi}^3}{32\pi F_\pi^2}
	-\frac{(g_1-3f_1)^2M_{\eta}^3}{96\pi F_{\eta}^2}\nonumber\\ 
	&&-\frac{(9f_1^2-6f_1g_1+5g_1^2)M_K^3}{48\pi F_K^2},		\\
	\nonumber\\
	\label{equ:21}
		m_{P_{\psi s}^{\Sigma}}&=&m_0-4b_{1}M_\pi^2	-\frac{g_1^2M_{\eta}^3}{24\pi F_{\eta}^2}\nonumber\\
	&&	-\frac{(6f_1^2+g_1^2)M_\pi ^3}{24\pi F_\pi^2}
	-\frac{(f_1^2+g_1^2) M_{K}^3}{8\pi F_K^2},
	\\
	\nonumber\\
	\label{equ:22}
		m_{P_{\psi s}^{\Lambda}}
	&=&m_0+\frac{4}{3}b_{1}(M_\pi^2-4M_K^2)\nonumber\\
	&&-\frac{g_1^2M_\pi ^3}{8\pi F_\pi^2}
	-\frac{(9f_1^2+g_1^2) M_{K}^3}{24\pi F_K^2}
	-\frac{g_1^2M_{\eta}^3}{24\pi F_{\eta}^2},
	\\
	\nonumber\\
	\label{equ:23}
		m_{P_{\psi ss}^{N}}&=&m_0+4b_{2}(M_\pi^2-M_K^2)-4b_{1}M_K^2\nonumber\\
	&&-\frac{3(f_1-g_1)^2 M_{\pi}^3}{32\pi F_\pi^2}
		-\frac{(3f_1+g_1)^2M_{\eta}^3}{96\pi F_{\eta}^2}\nonumber\\
	&&-\frac{(9f_1^2+6f_1g_1+5g_1^2)M_K^3}{48\pi F_K^2} .
\end{eqnarray}


For the physical masses of $8_1$ flavor hidden-charm pentaquark states, there are five LECs $m_0$, $b_1$, $b_2$, $f_1$, $g_1$. 
To estimate the LECs $g_{1}$ and $f_{1}$,  we consider the $\pi_{0}$ meson decay of $P_{\psi}^{N^{+}}$ and ${P_{\psi ss}^{N^{0}}}$ similar to the procedure employed for the nucleon in quark model. As shown in Ref.~\cite{Li:2024wxr}, we obtain $f_{1} = \frac{1}{3}g_{A}=0.42$, $g_{1} = \frac{1}{5}g_{A}=0.25$, $f_{4} =g_{4}=0$. 

In chiral perturbation theory, the remaining three LECs $m_0$, $b_1$, $b_2$ should be fitted through experimental inputs. Among the hidden-charm pentaquark states that have already been observed by LHCb Collaboration, $P_{\psi}(4312)$ is close to the $\bar{D}\Sigma_c$ and $\bar{D}^*\Sigma_c$ threshold whose spin-parity is $J^P=\frac{1}{2}^{-}$~\cite{Wang:2019nvm}, $P_{\psi{}s}^{\Lambda}(4338)$ is close to the $\Xi_c\bar{D}$ whose spin-parity is preferred to be $J^P=\frac{1}{2}^{-}$~\cite{LHCb:2022ogu}. We suppose $P_{\psi}(4312)$ and $P_{\psi{}s}^{\Lambda}(4338)$ are both $8_1$ flavor pentaquark states and take $m_{P_{\psi}^{N}}=4.312$GeV and $m_{P_{\psi s}^{\Lambda}}=4.338$GeV as inputs.

\begin{figure}[htbp]
			\centering
			\includegraphics[width=1.0\linewidth]{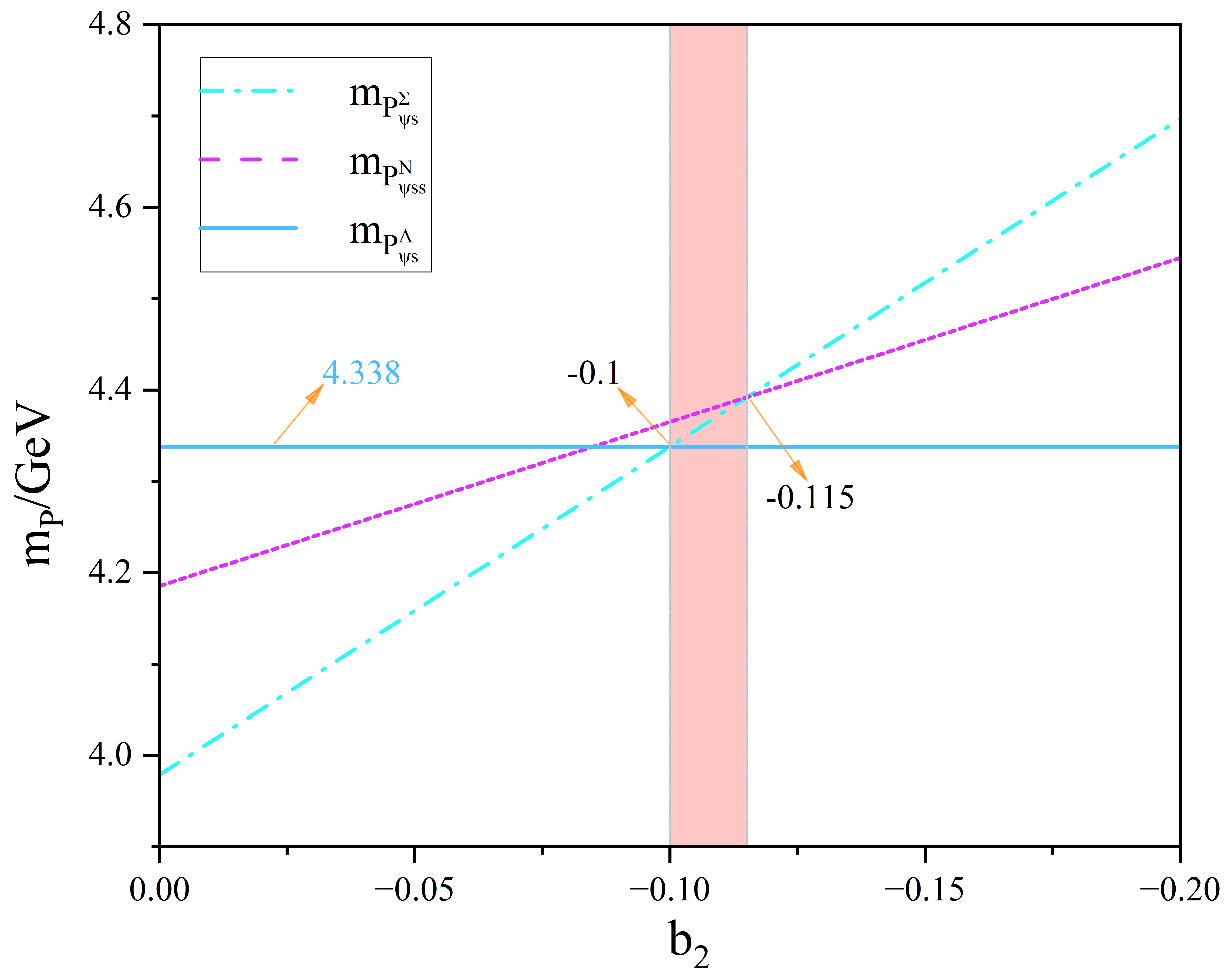}
			\caption{The variations of hidden-charm pentaquark mass with $b_2$. The pink band indicates the range of $b_2$.}
			\label{fig_2}
		\end{figure}

Further more, we obtain the remaining LEC through two simple constraint conditions.  
Since $\Sigma$ baryon is heavier than $\Lambda$ baryon considering chromomagnetic spin-spin interactions in the quark model, similarly, we have $m_{P_{\psi s}^{\Sigma}}$>$m_{P_{\psi s}^{\Lambda}}$. Since $s$ quark is heavier than $u$ quark and $d$ quark, we have $m_{P_{\psi s}^{\Sigma}}$<$m_{P_{\psi ss}^{N}}$. With these two constraint conditions, we show the variations of hidden-charm pentaquark masses with $b_2$ in Fig.~\ref{fig_2} and obtain
		\begin{eqnarray}
			-0.115<b_2<-0.100.
		\end{eqnarray}
		We take $b_2 =-0.108$, $b_1=-0.070$ and $m_0=4.466$GeV. Finally, we predict two other types of hidden-charm pentaquark states $P_{\psi s}^{\Sigma}$ and $P_{\psi ss}^{N}$,
		\begin{eqnarray}
			m_{P_{\psi s}^{\Sigma}}&=&4.367\mathrm{GeV},\\
			m_{P_{\psi ss}^{N}}&=&4.379\mathrm{GeV}.
		\end{eqnarray}
		\begin{figure}[htbp]
			\centering
			\includegraphics[width=1.0\linewidth]{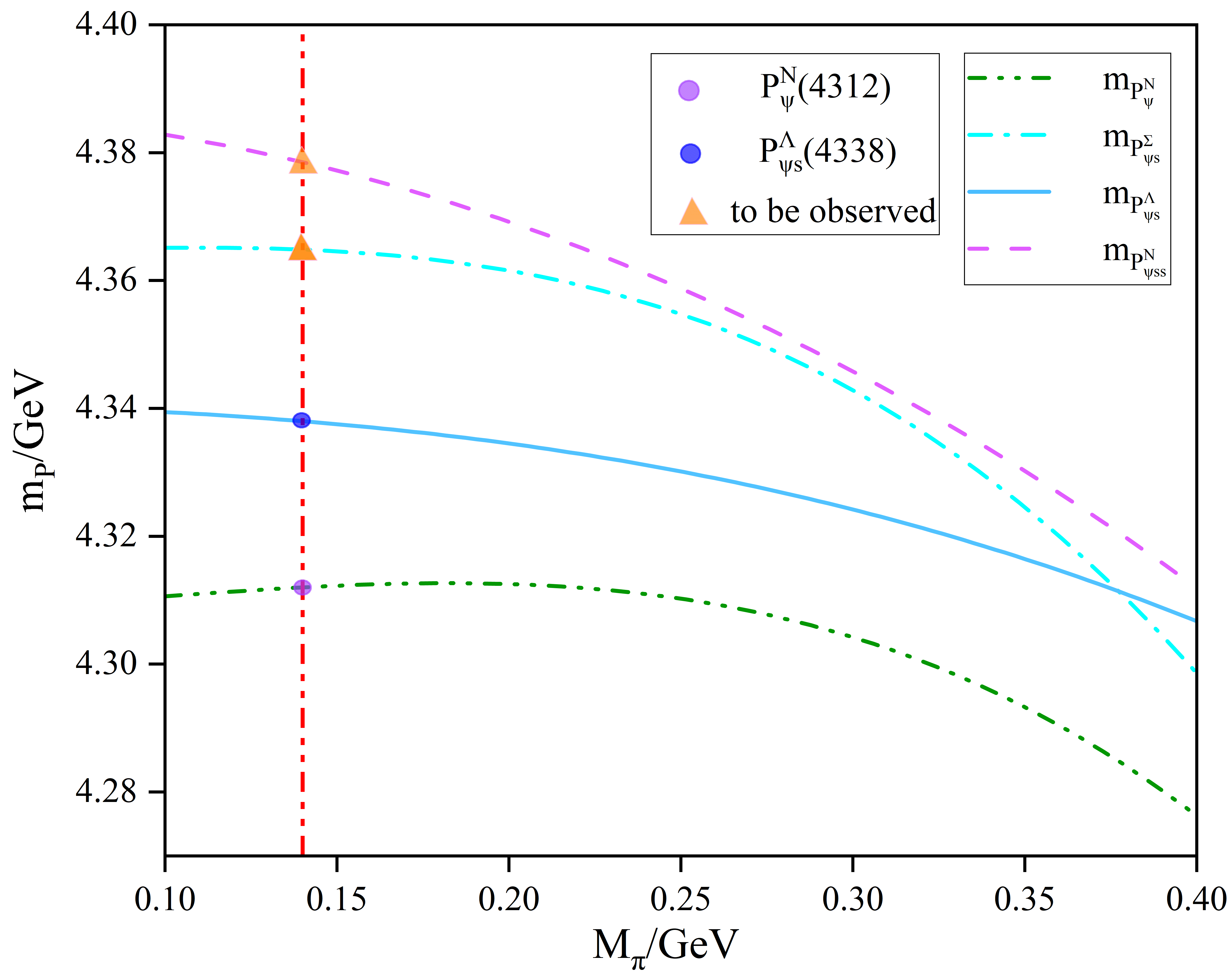}
			\caption{The variations of hidden-charm pentaquark masses with $M_\pi$. The red dashed line indicates the physical point where $M_\pi=0.140$GeV. The triangulation points represent the hidden-charm pentaquark states we predict}
			\label{fig_3}
		\end{figure}
	
\begin{figure}[htbp]
			\centering
			\includegraphics[width=1.0\linewidth]{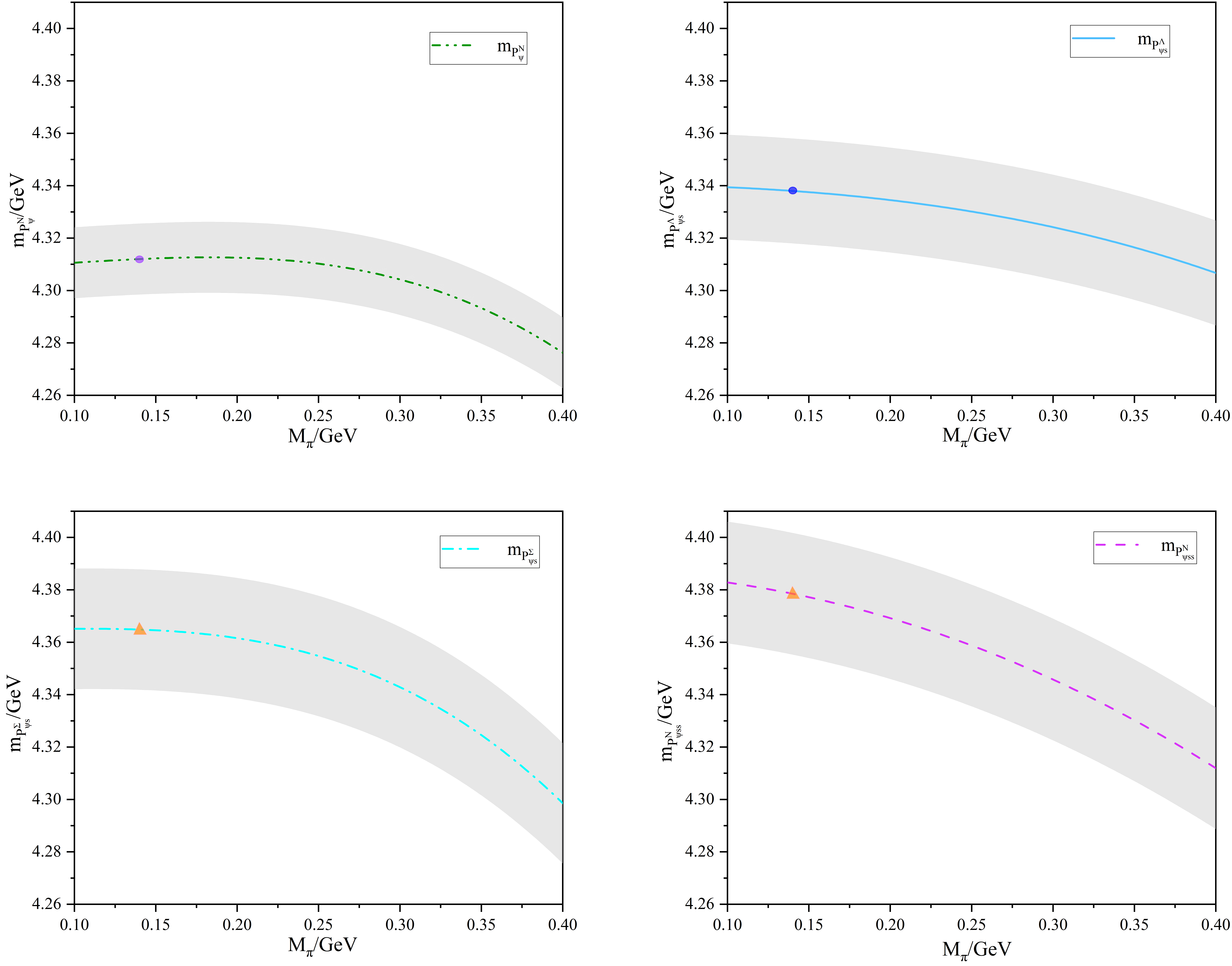}
			\caption{The hidden-charm pentaquark masses with uncertainties. The grey band depicts the uncertainty band of our determination.}
			\label{fig_4}
		\end{figure}

		In Fig.~\ref{fig_3}, we show the variations of hidden-charm pentaquark masses with $M_\pi$ and the triangulation points represent the hidden-charm pentaquark states we predict. It is evident that as $M_{\pi}$ increases, the pentaquark mass gradually decreases and $m_{P_{\psi s}^{\Sigma}}$ decreases faster than $m_{P_{\psi s}^{\Lambda}}$. When $M_\pi=0.378$GeV, masses of ${P_{\psi s}^{\Sigma}}$ and ${P_{\psi s}^{\Lambda}}$ are reversed. We suggest that lattice QCD pay attention to this interesting phenomenon in the future. Our analytical results will be useful to the possible chiral extrapolation of the lattice simulations. Realistic uncertainties of our calculation are determined in Fig.~\ref{fig_4}.
	  
	  If we suppose $P_{\psi}(4312)$ and $P_{\psi{}s}^{\Lambda}(4338)$ are both $8_2$ flavor pentaquark states, it is interesting to notice that the NLO loop contributions to the pentaquark masses are all zero due to $f_4=g_4=0$ in quark model. Similarly, we obtain $c_2 =-0.036$, $c_1=0.021$ and $m_0=4.365$GeV,	$m_{{P_2}_{\psi s}^{\Sigma}}=4.363\mathrm{GeV}$, $m_{{P_2}_{\psi ss}^{N}}=4.377\mathrm{GeV}$.

	   In short summary, we have investigated the octet hidden-charm molecular pentaquark masses to NLO and report on the first global study of octet molecular pentaquark masses up to one-loop order in the framework of HPChPT. Our calculations indicate that chiral perturbation theory can be performed to investigate the pentaquark states systematicaly and effectively. Our analytical results of the hidden-charm molecular pentaquark masses will be useful to the chiral extrapolation of lattice QCD.
	   
	   Since the LHCb Collaboration observed the first hidden-charm pentaquark states in 2015, with the continuous accumulation of experimental data, more and more pentaquark states have been observed except $P_{\psi s}^{\Sigma}$ and $P_{\psi ss}^{N}$. We predict $8_1$ flavor hidden-charm pentaquark states $P_{\psi s}^{\Sigma}(4367)$ and $P_{\psi ss}^{N}(4379)$ and $8_2$ flavor hidden-charm pentaquark states $P_{\psi s}^{\Sigma}(4363)$ and $P_{\psi ss}^{N}(4377)$. We suggest the LHCb Collaboration to observe $P_{\psi ss}^{N}(4379)^{-}$ and $P_{\psi ss}^{N}(4379)^0$ with $J^P=\frac{1}{2}^{-}$ in the $J/\psi \Xi$ spectrum through amplitude analyses of $\Omega_b^- \to J/\psi \Xi^0 K^-$ decays and $B^- \to J/\psi \Xi^- \bar{\Lambda}$ decays.

	  	  \section*{Acknowledgement}
	  This project is supported by the National Natural Science Foundation of China under Grants No. 11905171. This work is also supported by Shaanxi Fundamental Science Research Project for Mathematics and Physics (Grant No. 22JSQ016) and Young Talent Fund of Xi'an Association for Science and Technology (Grant No. 959202413087).

	\end{document}